# Multifunctional Magnetic Oxide-MoS$_2$ Heterostructures on Silicon

*Allen Jian Yang\*, Liang Wu, Yanran Liu, Xinyu Zhang, Kun Han, Ying Huang, Shengyao Li, Xian Jun Loh, Qiang Zhu, Rui Su, Ce-Wen Nan\*, Xiao Renshaw Wang\**

A. J. Yang, Y. Liu, S. Li, R. Su, X. Renshaw Wang
Division of Physics and Applied Physics, School of Physical and Mathematical Sciences, Nanyang Technological University, 637371, Singapore
E-mail: yangjian@ntu.edu.sg; renshaw@ntu.edu.sg

L. Wu, X. Zhang
Faculty of Material Science and Engineering, Kunming University of Science and Technology, Kunming, Yunnan, 650093, China

K. Han
Information Materials and Intelligent Sensing Laboratory of Anhui Province, Institutes of Physical Science and Information Technology, Anhui University, Hefei, 230601, China

Y. Huang
State Key Laboratory of Environment-friendly Energy Materials, Southwest University of Science and Technology, Mianyang, 621010, China

X. J. Loh, Q. Zhu
Institute of Materials Research and Engineering (IMRE), A*STAR, 2 Fusionopolis Way, Innovis, 138634, Singapore

Q. Zhu
School of Chemistry, Chemical Engineering and Biotechnology, Nanyang Technological University, 21 Nanyang Link, 637371, Singapore

R. Su, X. Renshaw Wang
School of Electrical and Electronic Engineering, Nanyang Technological University, 637371, Singapore

R. Su

MajuLab, International Joint Research Unit UMI 3654, CNRS, Université Côte d'Azur, Sorbonne Université, National University of Singapore, Nanyang Technological University, 637371, Singapore

C. W. Nan

State Key Laboratory of New Ceramics and Fine Processing, School of Materials Science and Engineering, Tsinghua University, Beijing, 100084, China

E-mail: cwnan@tsinghua.edu.cn

**Abstract**

Correlated oxides and related heterostructures are intriguing for developing future multifunctional devices by exploiting their exotic properties, but their integration with other materials, especially on Si-based platforms, is challenging. Here we demonstrate van der Waals heterostructures of $La_{0.7}Sr_{0.3}MnO_3$ (LSMO), a correlated manganite perovskite, and $MoS_2$ on Si substrates with multiple functions. To overcome the problems due to the incompatible growth process, technologies involving freestanding LSMO membranes and van der Waals force-mediated transfer were used to fabricate the LSMO-$MoS_2$ heterostructures. The LSMO-$MoS_2$ heterostructures exhibit a gate-tunable rectifying behaviour, based on which metal-semiconductor field-effect transistors (MESFETs) with on-off ratios of over $10^4$ can be achieved. The LSMO-$MoS_2$ heterostructures can function as photodiodes displaying considerable open-circuit voltages and photocurrents. In addition, the colossal magnetoresistance of LSMO endows the LSMO-$MoS_2$ heterostructures with an electrically tunable magnetoresponse at room temperature. Our work not only proves the applicability of the LSMO-$MoS_2$ heterostructure devices on Si-based platform but also demonstrates a paradigm to create multifunctional heterostructures from materials with disparate properties.

**1. Introduction**

As the physical dimension of transistors, the fundamental switches that perform the digital computation in ubiquitous silicon chips, approaches atomic scale, it has become increasingly difficult to improve the overall performance by merely shrinking the size of the transistors and increasing their density.[1] Monolithic integration of non-digital functions, including sensors, amplifiers, and actuators, with the Si platform has been explored as an alternative approach and shown great potential for future information-processing technologies.[2] This approach apparently requires heterogeneous integration of materials with appealing electronic, optical, magnetic, and mechanical properties on the Si platform without damaging the Si electronics. Two-dimensional layered materials (2DLMs) such as graphene and transition metal dichalcogenides (TMDCs), which have outstanding electronic and optical properties,[3] can be transferred from original growth substrates onto other platforms at low temperatures because they bond via weak van der Waals forces. These attributes make them especially useful for the heterogeneous integration applications.[4]

Due to their delicate interplays of charge, spin, orbital, and lattice, correlated oxides exhibit a variety of exotic properties, including colossal magnetoresistance (CMR), high-temperature superconductivity, and ferroelectricity.[5] Rejuvenated by the transformative achievements in the emerging 2DLMs, attempts to create heterostructures between correlated oxides and 2DLMs have been made by transferring 2DLMs onto the bulk oxide substrates or films epitaxially grown on lattice-matched oxide substrates. They have led to oxide-2DLM heterostructures on oxide substrates with many appealing functionalities, including voltage scaling of transistors, optoelectronic memories, redox-process-induced hysteresis of conductance, ferroelectric gating, and so on.[6] However, such a one-way transfer method fundamentally restricts the possible heterostructure configurations as well as their practical applications, because high-performance and low-cost electronic circuits require unlimited and monolithic integration with ubiquitous Si electronics.

Recently, freestanding crystalline perovskite oxides[7] have been realized as a detachable membrane by introducing a water-dissolvable sacrificial layer[8] or graphene interfacial layer[9] between the single-crystalline membrane and the oxide substrates. Transferring the freestanding oxide membranes onto desired materials or substrates could be a promising strategy to solve the aforementioned limitations of oxide-2DLM heterostructures. Here, we demonstrate multifunctional La$_{0.7}$Sr$_{0.3}$MnO$_3$ (LSMO)-MoS$_2$ heterostructures on Si by combining the freestanding LSMO membranes and deterministic transfer technologies for van der Waals heterostructures.[10] Different heterostructure configurations, manifested by the stacking order of materials, are readily achievable to accommodate the respective functionalities. Due to their distinct doping properties and energy band structures, the LSMO/MoS$_2$ heterostructures (LSMO on MoS$_2$) exhibit a gate-tunable rectifying behaviour, which is harnessed to fabricate metal-semiconductor field-effect transistors (MESFETs) with a current on-off ratio ($I_{on}/I_{off}$) of over $10^4$ and a subthreshold swing (SS) as low as 87 mV/dec. Based on the MoS$_2$/LSMO heterostructures (MoS$_2$ on LSMO) we demonstrate Schottky photodiodes which have a linear response to the power density of the illumination. The deliberately designed MoS$_2$/LSMO heterostructures also exhibit a rare magnetic response which can be controlled by the back-gate voltages.

## 2. Results and Discussion

### 2.1. Gate-tunable Schottky diodes

The LSMO-MoS$_2$ heterostructures were fabricated by sequentially assembling the component materials, i.e. freestanding LSMO and MoS$_2$. The step-by-step process of preparing freestanding LSMO flakes, transferring them onto the SiO$_2$/Si substrates, and fabricating LSMO-MoS$_2$ heterostructures is illustrated in Figure S1 (Supporting Information). First, the LSMO film grown on a SrTiO$_3$ (STO) substrate with a pre-grown Sr$_3$Al$_2$O$_6$ (SAO) sacrificial layer was released from the substrate by dissolving the middle SAO layer with deionized water.[8a] Then the LSMO film supported by a polydimethylsiloxane (PDMS) sheet, which weakly adhered to LSMO via van der Waals forces, could be transferred onto SiO$_2$/Si substrate due to a possibly stronger interaction between LSMO and SiO$_2$ than that between LSMO and PDMS.[10] The transferred LSMO flakes display characteristic electrical and magnetic properties as those of the epitaxial LSMO films,[11] manifested by the metallic conduction, ferromagnetism at room temperature, and colossal magnetoresistance (Figure S2, Supporting Information). When an LSMO flake was aligned and transferred on a MoS$_2$ flake, which had been prepared on the SiO$_2$/Si substrate by mechanical exfoliation, an LSMO-MoS$_2$ heterostructure was obtained and denoted as the LSMO/MoS$_2$ heterostructure as schematically shown in **Figure 1**a. Alternatively, when the transfer sequence was reversed, a MoS$_2$/LSMO heterostructure where MoS$_2$ was on top of LSMO was obtained. These two types of configurations were explored for different functionalities in this work.

The LSMO/MoS$_2$ heterostructures on SiO$_2$/Si exhibited a clear diode-like rectifying behaviour. Figure 1b shows the optical micrograph of an LSMO/MoS$_2$ heterostructure contacted by Au electrodes. The thicknesses of LSMO and MoS$_2$ are 43.2 nm and 7.9 nm, respectively, as measured by atomic force microscopy (AFM) (Figure 1c). Two-terminal current-voltage (*I-V*) measurements were carried out for different parts of the heterostructure. Symmetric *I-V* curves of individual LSMO and MoS$_2$ devices displaying considerable conductivity in Figure 1f indicate that the Au contacts to both materials are quasi-ohmic. It is worth noting that as a high-work-function metal, Au forms quasi-ohmic contact with n-doped MoS$_2$ because of the strong Fermi-level pinning at the Au-MoS$_2$ interface.[12] In contrast, *I-V* curves across the heterojunction are highly asymmetric and analogous to those of a conventional diode. The linear-scale *I-V* characteristics of the heterojunctions are shown in Figure S3, Supporting Information. The rectification ratio (RR), defined as the ratio of forward-biased to reverse-biased currents at the same magnitudes of voltage, is over 10$^3$ measured at ±2 V. The RR is limited by the levelled forward-biased current and considerable reverse-biased current. The former is partially due to the series resistance of LSMO and MoS$_2$ while the latter can be attributed to the small barrier height between LSMO and MoS$_2$, the surface-defect-related leakage path, or both. A vertical device structure comprising LSMO-MoS$_2$-metal stack may alleviate the series resistance problem and improve the forward-bias current. Surface modification of the LSMO with polar self-assembled monolayers (SAMs) could be a viable approach to increasing the barrier height and passivating the defects, thus reducing the reverse-biased current.[13] This rectifying characteristic can be used to build a half-wave rectifier as shown in Figure S4, Supporting Information. The negative half of the sinusoidal input voltage ($V_{IN}$) is blocked while the positive half is reproduced by the output voltage ($V_{OUT}$).

The rectifying characteristics of the LSMO/MoS$_2$ heterostructures can be attributed to the distinct doping properties of LSMO and MoS$_2$ and their energy band alignments. In addition to the monotonic decrease of resistivity with decreasing temperature (Figure S2a, Supporting Information), the fact that LSMO has no response to the back-gate voltage ($V_{BG}$) (Figure S5, Supporting Information) is also a reflection of its metallic (or degenerate doping) nature. In contrast, the MoS$_2$ device can be turned off by a negative $V_{BG}$, functioning as a typical n-type depletion-mode field-effect transistor (Figure S5, Supporting Information). This effect corroborates the intrinsic n-type doping of MoS$_2$ and is consistent with previous reports.[14] The n-type doping and considerable conductivity of MoS$_2$ in the absence of gate voltages are critical to the operation of the Schottky diodes and other devices described in the following sections. Other semiconducting TMDCs such as MoSe$_2$, MoTe$_2$, and WS$_2$ that are mostly insulating without chemical or electrostatic doping are not suitable for these device applications. The local work function difference across the heterojunction is revealed by mapping the contact potential difference (CPD) (Figure 1d) using scanning Kelvin probe microscopy (SKPM). The CPD reflects the difference between the work function of the SKPM tip ($\phi_{\text{tip}}$) and that of the sample ($\phi_{\text{LSMO}}$ and $\phi_{\text{MoS}_2}$) by $-eV_{\text{CPD}} = \phi_{\text{tip}} - \phi_{\text{LSMO}}$ and $-eV_{\text{CPD}} = \phi_{\text{tip}} - \phi_{\text{MoS}_2}$, where $e$ = 1.6 × 10$^{-19}$ C is the elementary charge. Considering that $\phi_{\text{tip}}$ is fixed, the work function difference between LSMO and MoS$_2$ can be deduced from the difference in $V_{\text{CPD}}$ ($\Delta V_{\text{CPD}}$) by $\Delta V_{\text{CPD}} = (\phi_{\text{LSMO}} - \phi_{\text{MoS}_2})/e$. According to the CPD profile in Figure 1d inset, LSMO has a work function of about 24.5 meV larger than that of MoS$_2$. Therefore, at thermal equilibrium there is a net electron transfer from MoS$_2$ to LSMO and consequently bending of the conduction and valence bands of MoS$_2$ as schematically shown in Figure 1e(i). As a result, the electrons need to overcome a significantly larger energy barrier to be thermionically emitted from LSMO to MoS$_2$ (Figure 1e(iii)) than the reverse (Figure 1e(ii)), thereby explaining the observed rectifying results.

The electrical characteristics of the LSMO/MoS$_2$ heterostructure device can be dramatically modulated by $V_{\text{BG}}$. As shown in Figure 1g, the rectifying behaviour diminishes at larger positive $V_{\text{BG}}$ because the reverse-biased current increases much faster with the $V_{\text{BG}}$ than the forward-biased current. Additionally, both forward and reverse currents vanish at large negative $V_{\text{BG}}$. These trends are shown more clearly by the transfer curves (Figure 1h) obtained by sweeping the $V_{\text{BG}}$ at a fixed bias. The RR has a peak value of 1.4 × 10$^4$ at around $V_{\text{BG}}$ = −15 V and is reduced to merely 7.7 at $V_{\text{BG}}$ = 40 V. When the bias is positive, the diode is forward biased and the junction resistance is relatively small. Thus, the heterostructure device behaves similarly to a depletion-mode MoS$_2$ transistor. On the contrary, at reverse biases the highly resistive LSMO-MoS$_2$ junction dominates the total resistance of the heterostructure device. The current through the device is low unless a large positive $V_{\text{BG}}$ is applied to electrostatically dope MoS$_2$. The electrostatic doping reduces the barrier thickness so that electrons from LSMO can tunnel through the barrier into MoS$_2$ (Figure 1e(iv)), dramatically increasing the reverse-biased current of the diode.[15] This gate tunability may be exploited to realize electronic and optoelectronic devices with desired behaviours.[16]

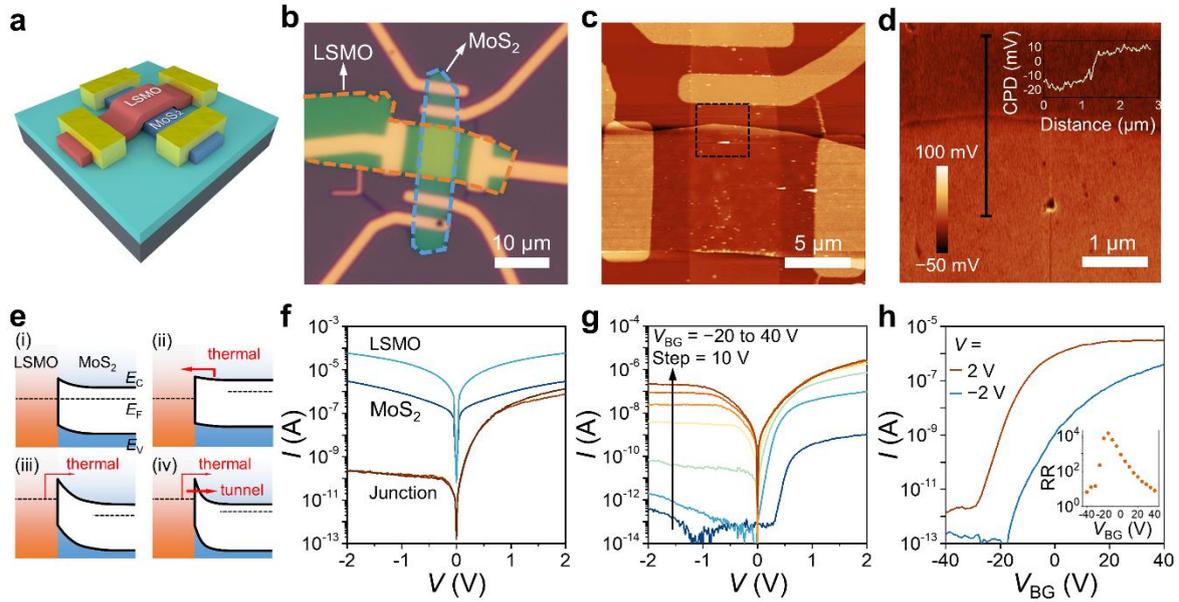

**Figure 1.** Electrical characteristics of the LSMO/MoS$_2$ heterostructure devices. a) Schematic of the device based on an LSMO/MoS$_2$ heterostructure. b) Optical micrograph of a typical LSMO/MoS$_2$ heterostructure device. The outlines of LSMO and MoS$_2$ flakes are denoted by the dashed lines of different colours. c) AFM image of the device shown in b). d) Contact potential difference (CPD) mapping of the region marked by the dashed rectangle in c). Inset, the profile of CPD along the dark line. e) Schematic energy band alignment and electron injection of the LSMO/MoS$_2$ heterostructure device (i) at thermal equilibrium, (ii) with a forward bias, (iii) with a reverse bias at zero $V_{BG}$, and (iv) with a reverse bias at a large positive $V_{BG}$. The red arrows represent the directions of electron flow. They are opposite to the electrical current directions. f) Semi-log plots of room-temperature current-voltage (*I*-*V*) characteristics measured on LSMO, MoS$_2$ and across the LSMO/MoS$_2$ junctions, respectively. g) *I*-*V* characteristics of the device at various $V_{BG}$, showing diminishing rectifying behaviour with increasing $V_{BG}$. h) Transfer curves (*I*-$V_{BG}$) of the device with forward and reverse biases. Inset, the rectification ratio (RR) at different $V_{BG}$.

## 2.2 Metal-semiconductor field-effect transistors (MESFETs)

The Schottky junction between LSMO and MoS$_2$ is harnessed to realize the operation of metal-semiconductor field-effect transistors (MESFETs). Similar to the combinations of a high-work-function metal, such as Pd and Pt, with an n-type semiconductor, including CdS,[17] ZnO,[18] MoS$_2$,[13b] etc., our MESFETs rely on controlling the conductance of the MoS$_2$ channel in proximity to LSMO with gate voltages ($V_G$) applied to LSMO. As depicted in **Figure 2**a, beneath LSMO there is a region in MoS$_2$ depleted of mobile electrons, hence insulating, due to the energy band alignment between LSMO and MoS$_2$ and their doping properties described in the previous section. When $V_G$ is negative, the LSMO-MoS$_2$ Schottky diodes are reverse biased and consequently the depletion region extends. In addition to being insulating, the depletion region also poses a significant energy barrier to the electrons flowing from the source (S) to the drain (D) electrode at positive drain-source voltages ($V_{DS}$). As a result, the drain-source current ($I_{DS}$) decreases with increasing negative $V_G$. At a sufficient negative $V_G$ the depletion region reaches the bottom of the MoS$_2$ channel and thus $I_{DS}$ diminishes. In other words, the transistor is at the off state. On the contrary, at positive $V_G$ the depletion region retracts towards LSMO. The bottom conductive layer allows the flow of electrons and thus $I_{DS}$ increases.

The electrical characteristics of our MESFETs are consistent with the operation mechanism described above. As shown in Figure 2b, at $V_{DS}$ of 1.0 V, $I_{DS}$ increases from $4.1 \times 10^{-11}$ to $1.4 \times 10^{-6}$ A when $V_G$ increases from −0.5 to 1.0 V, yielding an $I_{on}/I_{off}$ of $3.4 \times 10^4$. The minimum subthreshold swing (SS), the required $V_G$ span for a tenfold increase of $I_{DS}$ in the subthreshold region, is about 87 mV/dec, close to the thermionic limit of 60 mV/dec at room temperature. The transconductance $g$ of the MESFET, extracted via $g = \frac{dI_{DS}}{WV_{DS}dV_G}$ where $W$ is the channel width (6.4 μm), is shown in Figure S6. A peak transconductance of 0.32 μS μm$^{-1}$ is measured at $V_{DS}$ = 1.0 V. This value is limited by the relatively large channel length (9.3 μm), which is defined by the width of the LSMO flake, and low mobility of MoS$_2$. The conductance of the MoS$_2$ channel, which sets the upper limit of the conductance between the drain and source electrodes, is proportional to the mobility the MoS$_2$ and inversely proportional to the length of the channel. With proper patterning techniques to fabricate narrow LSMO strips, largely enhanced transconductance can be anticipated. Figure 2c shows the output characteristics of the MESFET. $I_{DS}$ increases linearly with $V_{DS}$ when $V_{DS}$ is small and saturates at larger $V_{DS}$ for all $V_G$ values. This is due to the expansion of the depletion region at the drain side that limits $I_{DS}$. Even though $V_G$ is positive, the gate-drain Schottky junction can be reverse-biased when $V_{DS}$ is large enough. Since the ferromagnetic Curie temperature ($T_c$) of the LSMO, which is around 310 K as revealed by the temperature-dependent magnetization (Figure S2c, Supporting Information), is only slightly above room temperature, the LSMO could undergo metal-insulator transition due to temperature fluctuation. This effect on the MESFET operation was investigated by heating the device above $T_c$ and measuring the transfer characteristics ($I_{DS}$-$V_G$). As shown in Figure S7 (Supporting Information), the MESFET displays similar transfer characteristics at room temperature (300 K) and at temperatures higher than $T_c$. Since the insulating/semiconducting phase of LSMO at temperatures above $T_c$ is highly conductive, it still forms a Schottky or p-n junction at the LSMO-MoS$_2$ interface. Thus, the aforementioned working mechanism of MESFETs is still effective.

The Si back gate can be used to modulate the electrical characteristics of the MESFET as shown in Figure 2d. A positive $V_{BG}$ accumulates more mobile electrons in the MoS$_2$, which increase the required negative $V_G$ to turn off the transistor. Thus, the $I_{DS}$-$V_G$ curves shift to the left with increasing $V_{BG}$. In addition, the increased electron concentration in MoS$_2$ reduces both the series resistance of the underlapping MoS$_2$ and contact resistance between MoS$_2$ and the Au electrodes. As a result, with increasing $V_{BG}$ the overall resistance decreases and $I_{DS}$ increases. In other words, the $I_{DS}$-$V_G$ curves shift upwards with increasing $V_{BG}$. These back-gate-tunable transfer characteristics make the MESFET adjustable for different applications. For example, it may be switched to a low-resistance mode for amplifiers, where a large $I_{DS}$ is required to boost operation speed whereas the power consumption due to $I_{off}$ is less important. Unlike metal-oxide-semiconductor field-effect transistors (MOSFETs), where the semiconducting channel is separated from the gate by dielectric materials, MESFETs, as well as junction field-effect transistors (JFETs) built on p-n junctions,[19] have a direct contact between the channel and the gate. Such configurations have the advantages of simpler structure and easier fabrication but at the cost of larger gate leakage current, especially when the Schottky or p-n junctions are forward-biased (Figure S8, Supporting Information). This gate leakage causes the deviation of $I_{DS}$-$V_G$ curves at small $V_{DS}$ from the expected monotonic increase in Figure 2b. It's also the main reason for the dramatically increased $I_{off}$ at large $V_{BG}$ in Figure 2d. In addition to the surface modification of LSMO with passivating SAMs as suggested in the previous section, integrating a thin insulating layer such as SrTiO$_3$ and LaAlO$_3$ between LSMO and MoS$_2$ to form a metal-insulator-semiconductor (MIS) tunnel junction could be beneficial to the mitigated gate leakage and improved power efficiency.[20]

The considerable $I_{on}/I_{off}$, small SS, and saturating $I_{DS}$ are beneficial for the low-voltage and high-gain operation of inverters, which are the basic building blocks for more complex logic and analog circuits. The resistive-load inverter, shown in Figure 2e, operates within a small input voltage ($V_{IN}$) range (1.0 V). At a supply voltage ($V_{DD}$) of 1.0 V the peak value of voltage gain, defined as ($-dV_{OUT}/dV_{IN}$) where $V_{OUT}$ is the output voltage, is 4.0 (Figure 2f), which is greater than unity. It implies that such inverters are suitable for building cascaded logic and analog circuits to process digital and analog signals. As shown in Figure 2g, the small input ($V_{IN}$) sinusoidal waveform is amplified by the inverter amplifier which outputs a waveform with a larger amplitude and negligible signal distortion. The dynamic voltage gain is greater than 1 up to 200 Hz and then decreases with growing frequency (Figure 2h).

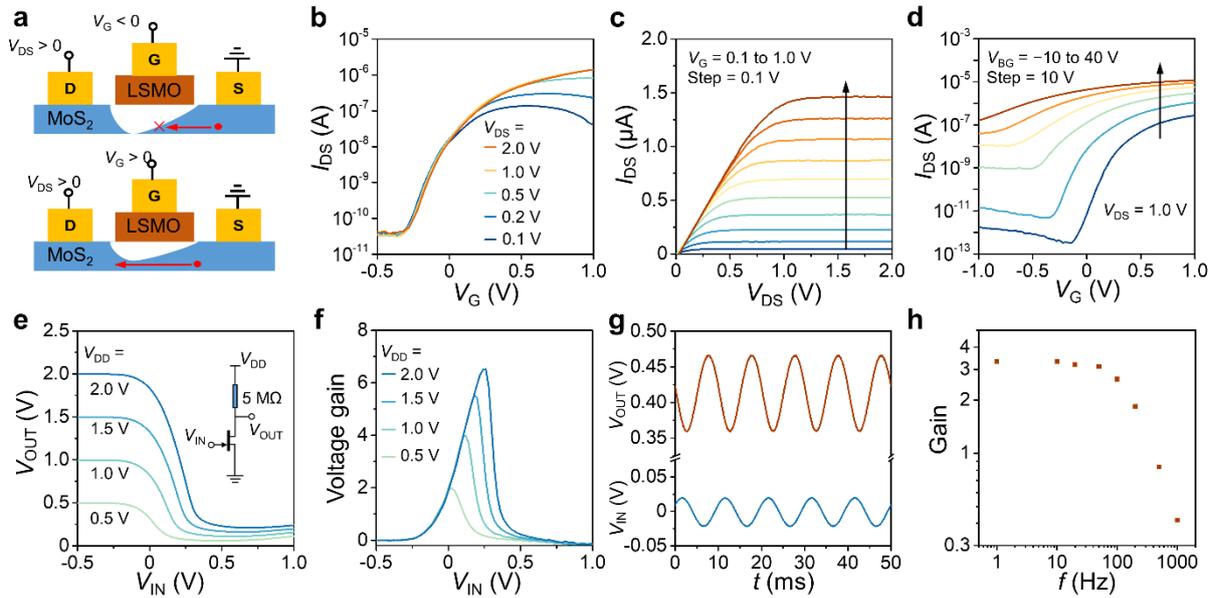

**Figure 2.** Metal-semiconductor field-effect transistors (MESFETs) relying on modulating the conductance of the $MoS_2$ channel by the LSMO gate. a) Cross-sectional schematic of the MESFET at the off (top) and on (bottom) states. The red dots and arrows represent electrons and the directions of electron flow, respectively. b) Transfer characteristics ($I_{DS}$-$V_G$) of the MESFET measured by sweeping the gate voltage ($V_G$) applied to LSMO while keeping the drain-source voltage ($V_{DS}$) fixed. c) Output characteristics ($I_{DS}$-$V_{DS}$) of the MESFET. d) Transfer characteristics ($I_{DS}$-$V_G$) of the MESFET at varying $V_{BG}$. e) Voltage-transfer characteristics of a resistive-load inverter built by connecting a resistor in series with the MESFET as shown in the inset. f) Voltage gains of the inverter in e), defined as ($-dV_{OUT}/dV_{IN}$), at different supply voltages. g) Amplification of the small input ($V_{IN}$) sinusoidal waveform (100 Hz) by the inverter amplifier. The amplitude of $V_{IN}$ waveform is 20 mV while that of the output ($V_{OUT}$) is 53 mV. h) Frequency ($f$) -dependent voltage gain of the amplifier.

## 2.3 Photodiodes based on MoS$_2$/LSMO heterostructures

The LSMO-MoS$_2$ heterostructures are explored as photodetecting devices in view of the outstanding optical properties of MoS$_2$. As depicted in **Figure 3**a, the photogenerated charge carriers in the depletion region can be separated by the built-in electric field at the junction. The negatively charged electrons flow to the MoS$_2$ side while the positively charged holes flow to the LSMO side. Collectively, such movements of charge carriers generate a current, i.e. photocurrent ($I_{ph}$), opposite to the direction of the Schottky diode, where LSMO is the anode and MoS$_2$ is the cathode. The total current flowing through the heterostructure is $I = I_{dark} + I_{ph}$, where $I_{dark}$ is the diode current at dark conditions. The semiconductor MoS$_2$ is on top of the metallic LSMO (Figure S9, Supporting Information) to facilitate absorption of the incoming photons by MoS$_2$. A relatively thick (66.5 nm) MoS$_2$ flake is used for this function because it absorbs more light but is not too thick to prevent the incoming photons from reaching the junction.[21] At dark conditions the MoS$_2$/LSMO heterostructure device exhibits a rectifying behaviour with a RR value of about $10^3$ at biases of ±1 V (Figure 3c). When the device is globally illuminated with a diffused 532 nm light, the reverse current increases significantly. In addition, a short-circuit current ($I_{sc}$), defined as the total current at zero bias, and an open circuit voltage ($V_{oc}$), measured by the applied bias at which the total current is zero, can be clearly observed. The $I_{sc}$ and $V_{oc}$ can also be clearly identified from the linear-scale $I$-$V$ curves under different illumination conditions (Figure S10). These characteristic photovoltaic effects indicate that the MoS$_2$/LSMO heterostructure device operates as a photodiode. To further verify this mechanism, we spatially map $I_{ph}$ through scanning a focused 532 nm laser (20 µW) over the device and simultaneously measuring $I_{ph}$. As shown in Figure 3b, $I_{ph}$, which is from MoS$_2$ to LSMO and therefore negative, is detected almost entirely from the overlapping region of MoS$_2$ and LSMO. In contrast, little $I_{ph}$ is detected in the sole LSMO or MoS$_2$ region. The spatial distribution and polarity of $I_{ph}$ confirm the built-in field at the junction is the origin of the photoresponse.

The dependence of the photoresponse of the Schottky photodiode on illumination power density ($P$) is also investigated. As shown in Figure 3d, $I_{ph}$ at $V = 0$ V, i.e. $I_{sc}$, increases monotonically with the power density. The fitting of $I_{ph}$-$P$ to power law relationship $I_{ph} = cP^{\alpha}$ reveals an $\alpha$ value of 1.06, which means $I_{sc}$ depends almost linearly on the power density. Similarly, $\alpha = 1.03$ is obtained for V = −0.5 V. The $V_{oc}$ has a roughly linear relationship with $\log_{10} P$ as shown in Figure 3d, consistent with the theoretical equation for $V_{oc}$ of photovoltaic devices.[22] The linearity of $I_{ph}$-$P$, along with considerable photocurrent, makes our photodiodes useful to work as photodetectors under both self-powered and conventional photoconductive modes. The $I_{ph}$ and $V_{oc}$ may be further improved by engineering the device structure and interfacial properties. A vertical graphene-MoS$_2$-LSMO structure, where the graphene acts as the transparent top electrode, may improve photocarrier collection efficiency.[23] The recombination of photogenerated charge carriers mediated by the defect states of LSMO could be alleviated by the aforementioned SAM and MIS approaches.

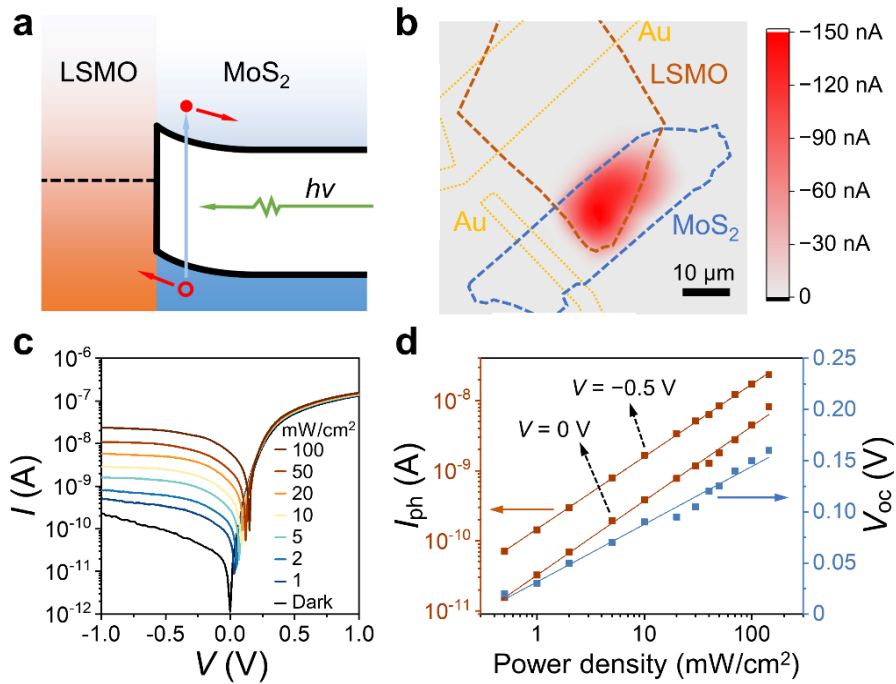

**Figure 3.** Photodiodes based on MoS$_2$/LSMO heterostructures. a) Energy band alignment and charge transport of the heterostructure under illumination. b) Photocurrent mapping of the heterostructure without external bias using a focused 532 nm laser with a power of 20 µW. c) Current-voltage ($I$-$V$) curves of the photodiode at dark conditions and under 532 nm laser illumination with power densities ranging from 1 to 100 mW/cm$^2$. d) Photocurrents ($I_{ph}$) and open-circuit voltages ($V_{oc}$) as functions of the illumination power density. The solid lines are the linear fittings of the corresponding experimental results.

## 2.4 Magnetoresponse of the MoS$_2$/LSMO heterostructure devices

The unique combination of materials in LSMO-MoS$_2$ heterostructures makes it possible to unite the functions of (i) diodes, (ii) field-effect transistors and (iii) magnetoresistors in one single device. To this end, we focus on MoS$_2$/LSMO heterostructures (MoS$_2$ on top of LSMO) (**Figure 4**a inset) where the electric field induced by the $V_{BG}$ is screened by the bottom LSMO. Thus the overlapped part of MoS$_2$ remains unaffected and the device is always rectifying regardless of the $V_{BG}$. In addition, a wide MoS$_2$ flake is used to match the conductance between LSMO and MoS$_2$ (Figure S11, Supporting Information). The current-voltage characteristics ($I_{DS}$-$V_{DS}$) in Figure 4a show a significant increase of $I_{DS}$ when the device is in a magnetic field of 8 T than in 0 T. The magnetocurrent ($I_B$) is extracted by $I_B = I(B) − I(0)$, where $I(B)$ and $I(0)$ are the current under a certain and zero magnetic field, respectively. Unlike the $I_B$ of a two-terminal LSMO device (Figure 4b), which increases almost linearly with the voltage magnitude for both forward and reverse biases, the $I_B$ of the heterostructure device can be switched on and off by a small bias. This controllability can be understood from the relative change between the heterojunction resistance and the series resistance of LSMO and MoS$_2$. With a sufficiently large forward bias, the Schottky diode is turned on, and its resistance is low compared to the series resistance of LSMO and MoS$_2$. Under this circumstance, the total resistance of the heterostructure device is dominated by LSMO if the series resistance of MoS$_2$ is also significantly smaller than that of LSMO. As a result, a significant $I_B$ emerges from the negative magnetoresistance of LSMO. On the contrary, the resistance of reverse-biased diode is orders of magnitude greater than the series resistance of LSMO and MoS$_2$. Thus, the heterostructure device shows no discernible $I_B$. The series resistance of MoS$_2$ is readily controlled by $V_{BG}$, thereby changing the overall magnetoresponse of the heterostructure devices. As shown in Figure 4c, $I_B$ is dramatically enhanced by the increased $V_{BG}$ when $V_{DS}$ is larger than 0.5 V. The heterostructure device can also work as a magneto-transistor of which transfer characteristics ($I_{DS}$-$V_{BG}$) depend on the strength of magnetic fields (Figure 4d). These attributes endow the heterostructure devices with more flexibility in potential sensing applications for high magnetic fields. Since the origin of the magnetoresponse of the heterostructure devices is the CMR of LSMO, improving the magnetoresistance of LSMO at room temperature is beneficial to enlarged magnetoresponse of the heterostructure devices. The matching of the resistances of the LSMO and MoS$_2$ flakes in the heterostructures also affects the overall magnetoresponse. These can be optimized by tuning the composition of LSMO, i.e. Sr-doping concentration, and growth conditions such as the temperature and oxygen partial pressure.[24]

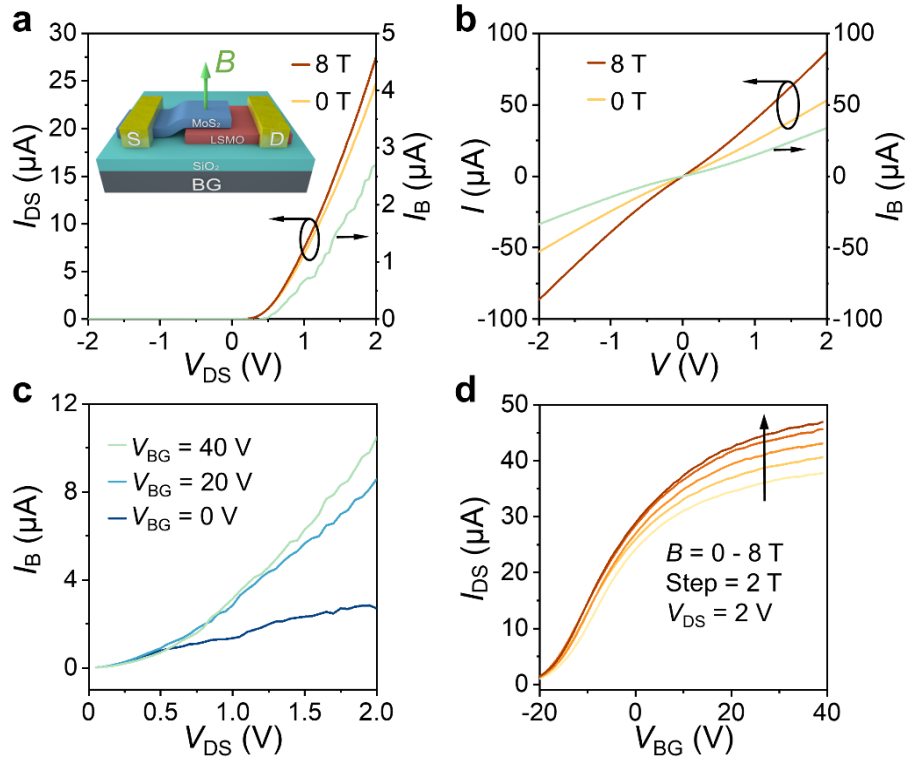

**Figure 4.** Magnetoresponse of the MoS$_2$/LSMO heterostructure devices. a) Current-voltage characteristics ($I_{DS}$-$V_{DS}$) of the heterostructure device under magnetic fields of 0 and 8 tesla at 300 K, and the corresponding magnetocurrent ($I_B$). Inset, schematic of the heterostructure device. A magnetic field is applied vertically to the heterostructure plane. b) Magnetoresponse of a two-terminal LSMO device. c) The dependence of $I_B$ on $V_{DS}$ and $V_{BG}$. d) Transfer curves ($I_{DS}$-$V_{BG}$) of the heterostructure device with fixed $V_{DS}$ measured under different levels of magnetic fields.

## 3. Conclusion

In summary, we developed multifunctional LSMO-MoS$_2$ heterostructures on Si substrates manifested by Schottky diodes, MESFETs, photodiodes and magnetoresponsive diodes and transistors. These devices may be integrated as rectifiers, high magnetic field sensors, and amplifiers in Si integrated circuits. Combining the scaled-up growth and transfer of large-area LSMO and MoS$_2$ and the recently proven mechanical flexibilities of these materials, these heterostructure devices have a potential for large-area low-power flexible electronics and optoelectronics.[8d, 25] In addition, our construction scheme can be generalized to heterostructures based on a vastly wide range of correlated oxides and 2DLMs. It also can be used to create more complex heterostructures with multiple layers. Unlimited combinations are highly beneficial to the exploration of unusual physical properties and a broad range of applications.

**4. Experimental Section**

*Material preparation*: SAO and LSMO were sequentially grown on STO (001) substrates using pulsed laser deposition (PLD) in the oxygen partial pressures ($P_{O2}$) of 1 x 10$^{-6}$ Torr and 0.2 Torr, respectively. All STO substrates were TiO$_2$-terminated, which were achieved by aqua regia etching and then 950 °C annealing in air. The targets were sintered polycrystalline SAO and LSMO pallets, and were ablated by an excimer laser ($\lambda$=248 nm) with a repetition rate of 2 Hz and a laser fluence of 1.8 J/cm$^2$. During the growth, the STO substrates were kept at 760 °C. All samples were cooled down to room temperature in the growth pressure of LSMO, 0.2 Torr. MoS$_2$ flakes were mechanically exfoliated from a bulk crystal and transferred onto SiO$_2$/Si. The Si substrate (~0.675 mm thick) was p-doped with a resistivity smaller than 1 ohm-cm and covered with 285 nm SiO$_2$. When transferring MoS$_2$ onto LSMO was needed, the MoS$_2$ flakes on SiO$_2$/Si were first transferred onto a polydimethylsiloxane (PDMS) sheet using a wedging transfer method.[26]

*Material characterizations*: The magnetic moment of the as-grown and transferred LSMO films were tested with a Magnetic Property Measuring System (MPMS) (Quantum Design Inc.). The topography and SKPM of MoS$_2$, LSMO, and the heterostructures on SiO$_2$/Si were measured with a Cypher ES Environmental atomic force microscope (Asylum Research).

*Device fabrication*: Electrical contacts were fabricated by standard electron-beam lithography followed by lift-off of thermally evaporated Au (40-80 nm).

*Electrical measurements*: The devices were wire-bonded to chip-carriers for electrical measurements except the electrical measurements on the heating stage where the devices were directly probed in air using micromanipulators. The SiO$_2$ was scratched by a diamond pen to allow bonding from the conductive Si substrate which acted as the back gate. Magnetotransport studies were carried out in a cryogenic station (Oxford Instruments) equipped with a superconducting magnet. The Keithley 6221 sourcemeter and Keithley 2182 nanovoltameter were utilized for the four-terminal AC measurement. The electrical measurements at room temperature and without magnetic field were carried out in a shielded vacuum chamber with a pressure < 0.1 Torr. The Keithley 2635b and 2450 sourcemeters, and Keithley 4200A-SCS parameter analyzer were used for the DC measurements. The AFG1022 function generator (Tektronix) and TBS1052B oscilloscope (Tektronix) were used for the dynamic measurements on the rectifier and amplifier circuits. Photoelectric measurements were carried out in a scanning photocurrent system (Tuotuo Technology) at ambient conditions.

**Supporting Information**

Supporting Information is available from the Wiley Online Library or from the author.

**Acknowledgements**

A.J.Y. and L.W. contributed equally to this work. We thank the support and useful discussion from S.W. Zeng, D.Y. Wan, and Ariando. X.R.W. acknowledges supports from Academic Research Fund (AcRF) Tier 2 (Grant No. MOE-T2EP50120-006 and MOE-T2EP50220-0005) from Singapore Ministry of Education and Agency for Science, Technology and Research (A*STAR) under its AME IRG grant (Project No. A20E5c0094). R.S. gratefully acknowledges the funding support from Nanyang Technological University via Nanyang Assistant Professorship Start Up Grant. X.R.W. and R.S. acknowledge support from the Singapore Ministry of Education via the AcRF Tier 3 programme "Geometrical Quantum Materials" (Grant No. MOE2018-T3-1-002). L. W. acknowledges support from Natural Science Foundation of China (Grant No. 52102131) and Yunnan Fundamental Research Projects (Grant No. 202101BE070001-012 and 202201AT070171).

**Multifunctional Magnetic Oxide-MoS$_2$ Heterostructures on Silicon**

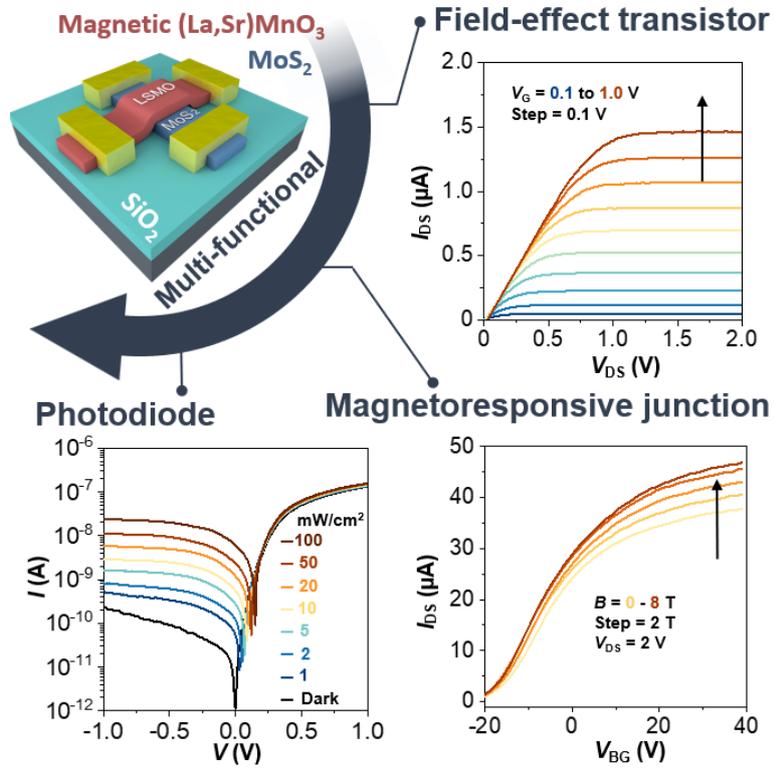

# Supporting Information

**Multifunctional Magnetic Oxide-MoS₂ Heterostructures on Silicon**

*Allen Jian Yang*, Liang Wu, Yanran Liu, Xinyu Zhang, Kun Han, Ying Huang, Shengyao Li, Xian Jun Loh, Qiang Zhu, Rui Su, Ce-Wen Nan*, Xiao Renshaw Wang**

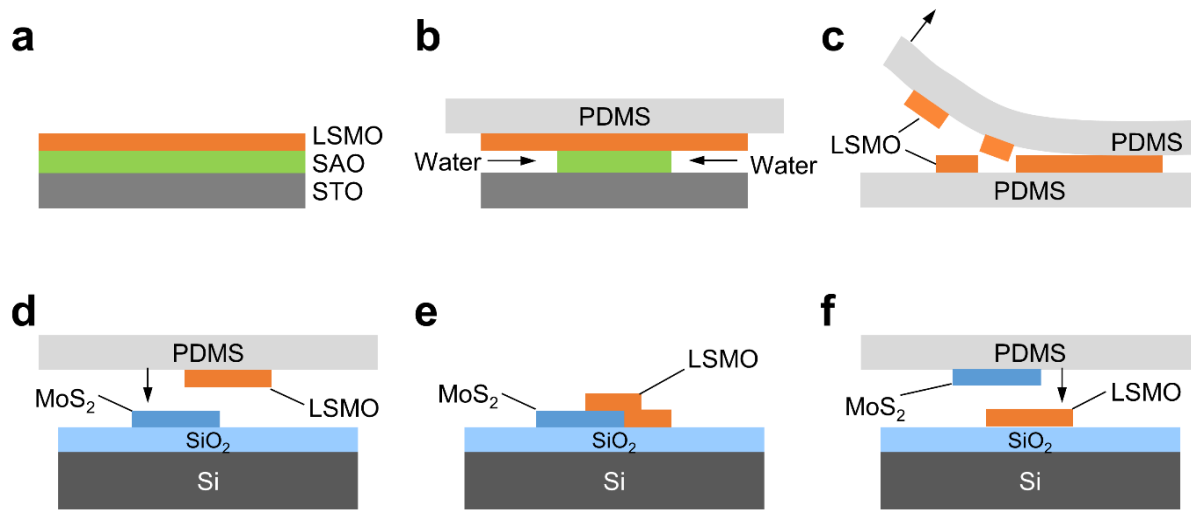

**Figure S1.** Schematic illustration of the process of producing freestanding LSMO flakes and La$_{0.7}$Sr$_{0.3}$MnO$_3$ (LSMO)-MoS$_2$ heterostructures. (a) Sequential growth of Sr$_3$Al$_2$O$_6$ (SAO) and LSMO on the SrTiO$_3$ (STO) substrate. (b) The surface of the LSMO/SAO bilayer was covered with a Polydimethylsiloxane (PDMS) sheet. Then, by immersing the stack in deionized (DI) water to dissolve the sacrificial SAO layer, the LSMO membrane supported by the PDMS floated above the water. (c) Laminating with another PDMS sheet and peeling yielded micrometre-sized LSMO flakes on the PDMS. (d) With the help of a microscope, the LSMO fakes were aligned and laminated onto a MoS$_2$ flake on SiO$_2$/Si. (e) The PDMS sheet was slowly peeled off, leaving LSMO on the SiO$_2$/Si surface. (f) The MoS$_2$/LSMO heterostructures were fabricated by aligning and laminating MoS$_2$ from the PDMS onto LSMO that had been transferred on SiO$_2$/Si.

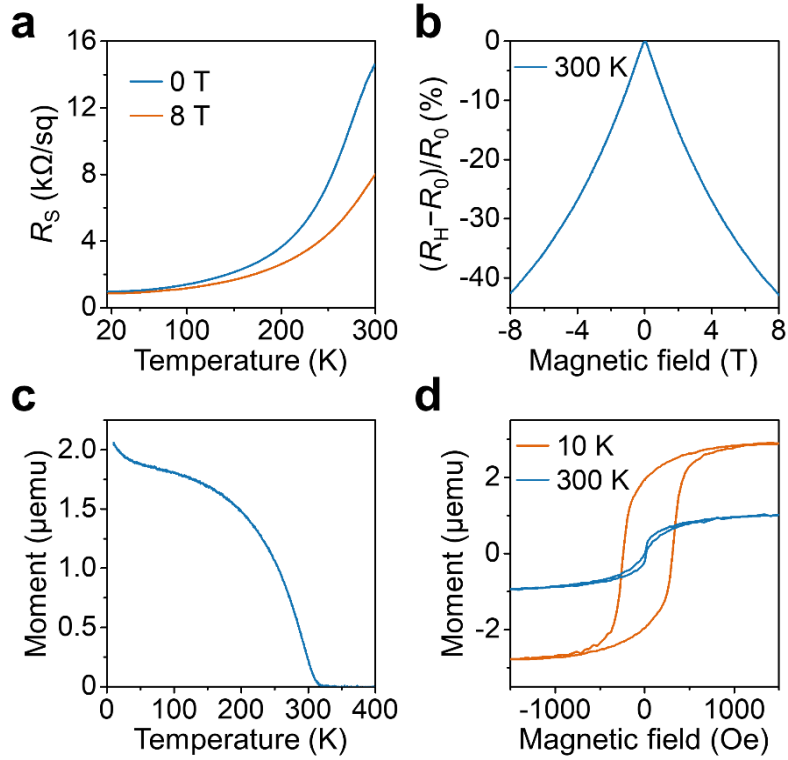

**Figure S2.** Electrical and magnetic properties of LSMO on $SiO_2$/Si. (a) Temperature-dependent sheet resistance ($R_S$) of the LSMO flake on $SiO_2$/Si under 0 and 8 T magnetic fields. (b) Magnetoresistance, defined as ($R_H$-$R_0$)/$R_0$, at 300 K as a function of the magnetic field. $R_H$ and $R_0$ are the $R_S$ of LSMO under finite and zero magnetic field, respectively. (c) Magnetization of a collection of LSMO flakes on $SiO_2$/Si as a function of temperature. (d) Magnetization hysteresis curves measured at 10 and 300 K.

The magnetoresistance of the LSMO flakes transferred on $SiO_2$/Si was investigated by measuring isolated LSMO flakes using the four-terminal contact scheme. Such devices were fabricated using similar methods as the LSMO-$MoS_2$ heterostructure devices which involved eletron-beam lithography and thermal deposition of Au. As shown in Figure S2a, the sheet resistance ($R_S$) of the LSMO flake (40.8 nm thick measured by AFM) decreases monotonically with decreasing temperature, indicating the metallic conduction of LSMO. Furthermore, in the presence of a magnetic field, $R_S$ decreases dramatically at all temperatures. Figure S2b shows the four-terminal room-temperature magnetoresistance (MR), which is defined as ($R_H$-$R_0$)/$R_0$, of the LSMO flake as a function of the magnetic field, where $R_H$ and $R_0$ are the $R_S$ at a definite and zero magnetic field, respectively. The magnitude of the negative MR increases sharply as the magnetic field increases and reaches over 40% at a magnetic field of 8 T. These properties are similar to LSMO films grown on oxide substrates, which are a p-type magnetic material with a unique CMR effect at room temperature.[1-2] This similarity indicates that our transferring process does not deteriorate the electrical properties of LSMO flakes. The temperature-dependent magnetization (Figure S2c), which was measured on all the flakes on the same $SiO_2$/Si, is consistent with the trend of $R_S$. It displays an abrupt increase at around 310 K, which marks the Curie temperature ($T_c$) of LSMO, affirming room-temperature ferromagnetism. This magnetic property is corroborated by the magnetization-field scan as shown in Figure S2d. The hysteresis loops clearly prove the ferromagnetism of LSMO transferred on $SiO_2$/Si substrates.

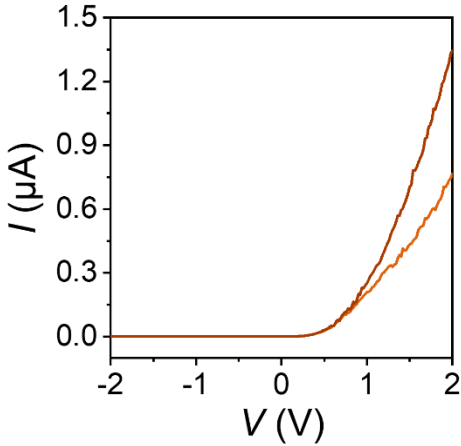

**Figure S3**. Linear-scale *I-V* characteristics of the heterojunctions in Figure 2.

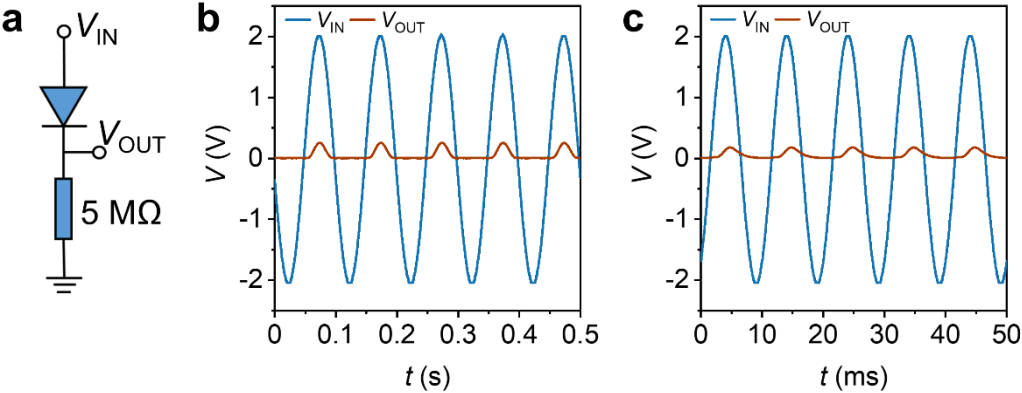

**Figure S4.** a) The circuit diagram of the half-wave rectifier built on the LSMO-$MoS_2$ Schottky diode. Output voltage ($V_{OUT}$) waveform in response to the input sine waveform ($V_{OUT}$) at b) 10 Hz and c) 100 Hz.

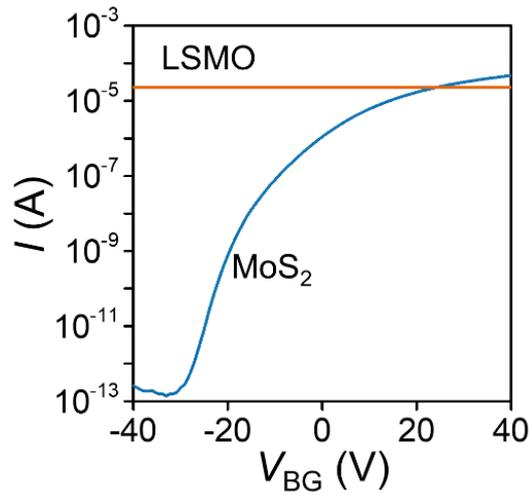

**Figure S5.** Back-gate transfer characteristics of LSMO and MoS$_2$ devices both on SiO$_2$/Si and at $V$ = 1 V. The $V_{BG}$ was applied to the p-doped Si substrate.

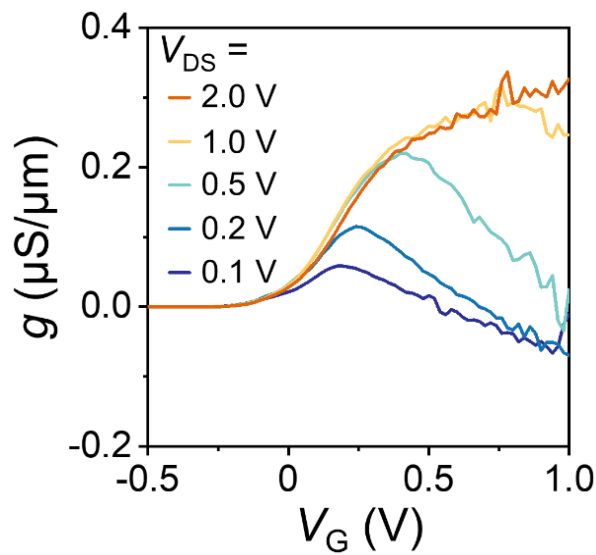

**Figure S6.** Transconductance ($g$) of the MESFET shown in Figure 2. At $V_{DS}$ = 1.0 V, the MESFET exhibits a peak transconductance of 0.32 µS µm$^{-1}$. Further increasing $V_{DS}$ does not increase $g$ because the transistor enters the saturation region, as revealed by the $I_{DS}$-$V_{DS}$ characteristics in Figure 2c.

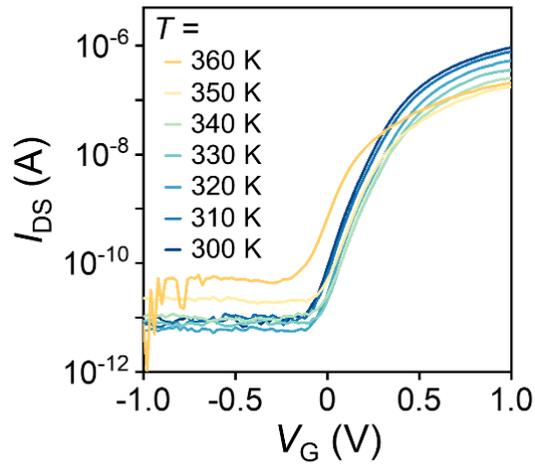

**Figure S7**. Transfer characteristics ($I_{DS}$-$V_G$) of the MESFET at $V_{DS}$ = 1.0 V measured at different temperatures in air. The MoS$_2$ channel can still be effectively switched on and off by applying positive and negative gate voltages ($V_G$) to LSMO, respectively, at temperatures higher than $T_c$.

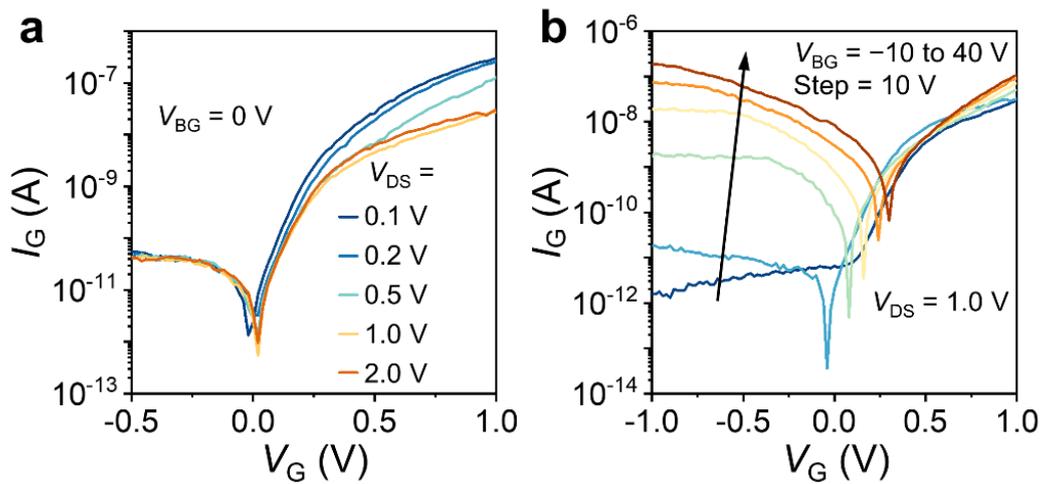

**Figure S8.** Gate leakage current ($I_G$) of the MESFET (a) at different $V_{DS}$ and (b) $V_{BG}$.

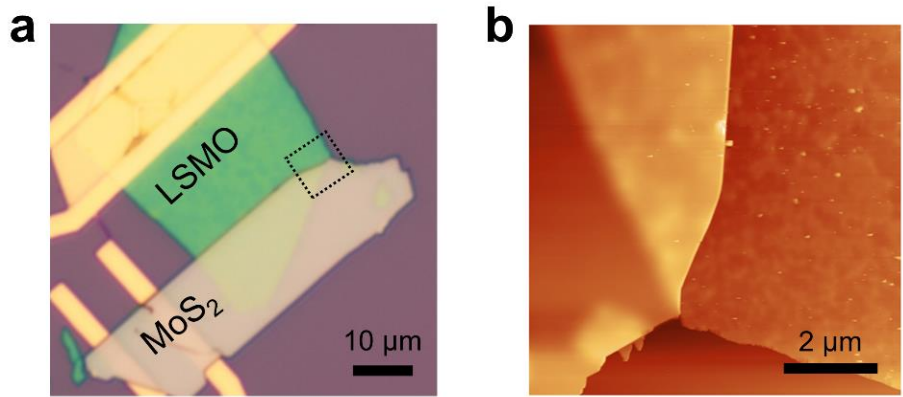

**Figure S9.** (a) Optical image of the MoS$_2$/LSMO heterostructure. (b) AFM image of the region marked by the dashed rectangle in (a).

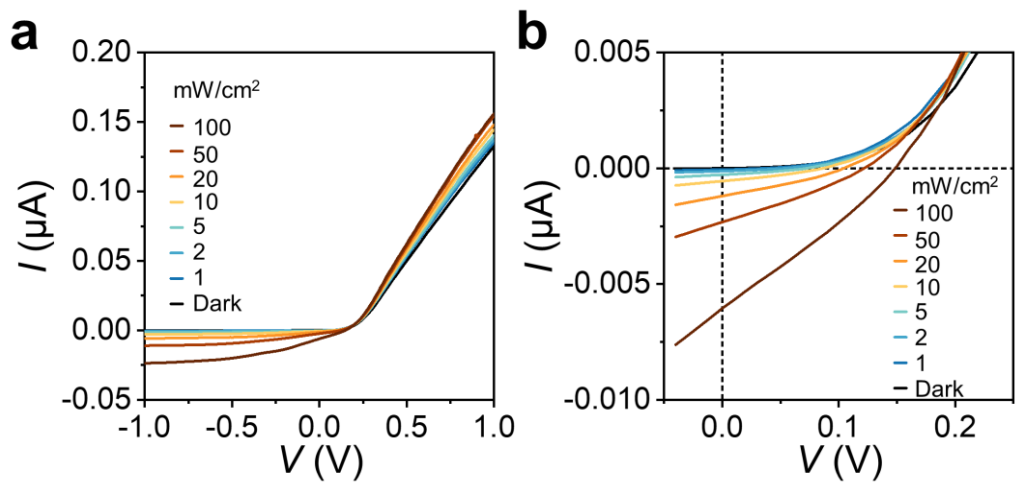

**Figure S10.** *I-V* curves in a linear scale of the photodiode under different illumination conditions shown in Figure 3. Panel b is the zoomed-in graph of panel a. The short-circuit current ($I_{sc}$) can be extracted from the intersections between the *I-V* curves and the vertical dashed line ($V = 0$ V), and the open-circuit voltage ($V_{oc}$) from intersections between the *I-V* curves and the horizontal dashed line ($I = 0$ A).

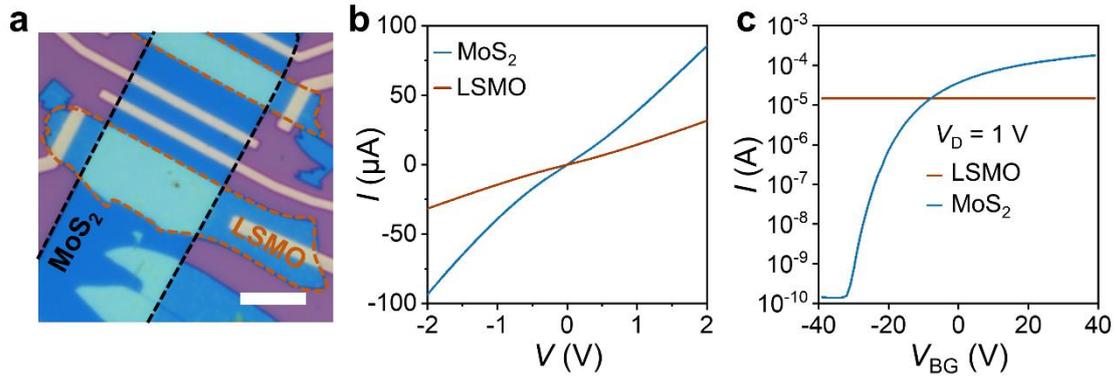

**Figure S11.** (a) Optical image of the MoS$_2$/LSMO heterostructure with a wide MoS$_2$ flake. Scale bar, 15 µm. (b) The *I-V* curves of the sole MoS$_2$ and LSMO show a larger current of MoS$_2$ compared to LSMO under zero $V_{BG}$. (c) Transfer characteristics (*I-$V_{BG}$*) of the individual LSMO and MoS$_2$ device. The MoS$_2$ device has a typical n-type field-effect transistor characteristic while the LSMO device has little response to $V_{BG}$. In addition, there is a significant crossover between the *I-$V_{BG}$* curves, which means the conductance of MoS$_2$ can be tuned much higher or lower than that of LSMO by $V_{BG}$.